\documentclass[proof]{WileyASNA-v1}
\usepackage{siunitx}

\articletype{Article Type}%

\received{14 January 2019}
\accepted{27 January 2019}

\begin{document}

\title{The short wavelength instrument for the polarization explorer balloon-borne experiment: Polarization modulation issues}

\author[1,2]{Fabio* Columbro }

\author[1,2]{Elia Stefano Battistelli }

\author[1,2]{Alessandro Coppolecchia }

\author[1,2]{Giuseppe D'Alessandro }

\author[1,2]{Paolo de Bernardis }

\author[1,2]{Luca Lamagna }

\author[1,2]{Silvia Masi }

\author[3,4]{Luca Pagano }

\author[1,2]{Alessandro Paiella }

\author[1,2]{Francesco Piacentini }

\author[1]{Giuseppe Presta }

\authormark{COLUMBRO F. \textsc{et al}}

\address[1]{\orgdiv{Dipartimento di Fisica}, \orgname{Sapienza Universit\'a di Roma}, \orgaddress{\country{Italy}}}

\address[2]{\orgdiv{INFN}, \orgname{sezione di Roma}, \orgaddress{\country{Italy}}}

\address[3]{\orgdiv{Dipartimento di Fisica e Scienze della Terra}, \orgname{Universit\'a degli studi di Ferrara}, \orgaddress{\country{Italy}}}

\address[4]{\orgdiv{INFN}, \orgname{sezione di Ferrara}, \orgaddress{\country{Italy}}}

\corres{*Corresponding author \email{fabio.columbro@roma1.infn.it}}

\presentaddress{Dipartimento di Fisica, Sapienza Universit\'a di Roma, 00185 Roma, Italy}

\abstract{In this paper we investigate the impact of using a polarization modulator in the Short Wavelenght Instrument for the Polarization Explorer (SWIPE) of the Large Scale Polarization Explorer (LSPE). The experiment is optimized to measure the linear polarization of the Cosmic Microwave Background at large angular scales during a circumpolar long-duration stratospheric balloon mission, and uses multi-mode bolometers cooled at \SI{0.3}{\kelvin}. The 330 detectors cover 3 bands at \SI{140}{\giga\hertz}, \SI{220}{\giga\hertz} and \SI{240}{\giga\hertz}.

Polarimetry is achieved by means of a large rotating half-wave plate (HWP) and a single wire-grid polarizer in front of the arrays. The polarization modulator is the first polarization-active component of the optical chain, reducing significantly the effect of instrumental polarization. A trade-off study comparing stepped vs spinning HWPs drives the choice towards the second. Modulating the CMB polarization signal at 4 times the spin frequency moves it away from $1/f$ noise from the detectors and the residual atmosphere.

The HWP is cooled at \SI{1.6}{\kelvin} to reduce the background on the detectors. Furthermore its polarized emission combined with the emission of the polarizer produces spurious signals modulated at $2f$ and $4f$. The $4f$ component is synchronous with the signal of interest and has to characterized to be removed from cosmological data.}


\maketitle

\section{Introduction}\label{sec1}
One of the most interesting goals in Cosmology nowadays is the detection of the primordial curl component (B-mode) of the Cosmic Microwave Background (CMB) polarization. A detection of B-modes at large angular scales would strongly reinforce the  evidence that cosmological inflation occurred (see e.g. \cite{BaumannPeiris:article}, since this peculiar polarization pattern is produced by gravitational waves generated by inflation in the early universe.

Detection of this tiny signal requires wide field of view mm-wave polarimeters, with large format arrays of detectors. Technologies replicating a single pixel in a very large low-power array, compatible with the cooling power of current cryogenic systems, are transition edge sensor (TES) bolometers (see e.g. \cite{Ade:article}) and kinetic inductance detectors (KIDs) (see e.g. \cite{McCarrick:article, Paiella_2019}).

The magnitude of the gravitational-wave B-mode signal is known to be less than \SI{1}{\micro\kelvin} {\it rms}. Detecting such a small signal from the ground or from a stratospheric balloon is a challenge, due to variable atmospheric transmission and noise. $1/f$ noise is particularly troublesome. For this reason, rapid modulation of linear polarization by a half-wave plate (HWP) is one of the most promising modulation techniques to reduce the effect of atmospheric and instrument fluctuations. This is especially true for instruments with large optical throughput and large numbers of pixels.
One of the main advantages of this approach consists in the rotation of the polarized component of the radiation at twice the rotation frequency, which is picked up in the detector timestreams modulated at four times the mechanical rotation frequency. On the other hand most of the spurious signals~\citep{Essinger:article} are modulated at $1f$, $2f$ and $4f$, so part of the systematic effects are efficiently removed by spectral filtering. Another advantage of using the HWP consists in the ability to make the polarimeter broadband, with a multi-plate HWP design \citep{Pisano:sapphire} or with optimized metamaterial meshes \citep{Pisano:mesh}.

Thanks to its continuous polarization modulator the long-duration balloon SWIPE-LSPE~\citep{SWIPE:article} will map the CMB polarization at large angular scales, with a wide sky coverage ($\sim 25\%$), to improve the limit on the ratio of tensor to scalar perturbations amplitudes down to $r = 0.01$.

After a brief overview of SWIPE instrument, in this paper we motivate the choice of the polarization modulation strategy and describe its implementation. Due to the requirements of cryogenic temperature and continuous rotation, we selected a superconducting magnetic bearing (SMB) \citep{Matsumura:article,Johnson:article} as the technology to spin the HWP and to modulate the polarized signal at a sufficiently high frequency (order of few Hz). An innovative frictionless clamp/release device \citep{Actuator:article}, based on electromagnetic actuators, keeps the rotor in position at room temperature, and releases it below the superconductive transition temperature, when magnetic levitation works properly.



\section{SWIPE}\label{sec2}
The Large-Scale Polarization Explorer (LSPE) \citep{LSPE:proceeding} is an experiment designed to measure the polarization of the CMB at large angular scales and to constrain the B-modes produced by tensor perturbations. LSPE is a joint experiment: STRIP (low-frequency ground-based telescope) and SWIPE (high-frequency balloon-borne telescope) which will provide high-precision complementary measurements of CMB polarization in the range  $44-240$ \SI{}{\giga\hertz}. 

SWIPE is a balloon-borne telescope flown on a stratospheric balloon, above most of the Earth atmosphere. The payload will fly in a circumpolar long duration mission ($\sim$ 15 days long) during the polar night. The balloon will be launched from the Longyearbyen airport (Svalbard Islands, \SI{78}{\degree}N latitude). The instrument will spin in azimuth, observing a large fraction of the northern sky ($\sim 25\%$) using the Earth as a giant solar shield.

\begin{figure}[ht]
\centering
{\includegraphics[scale=0.6]{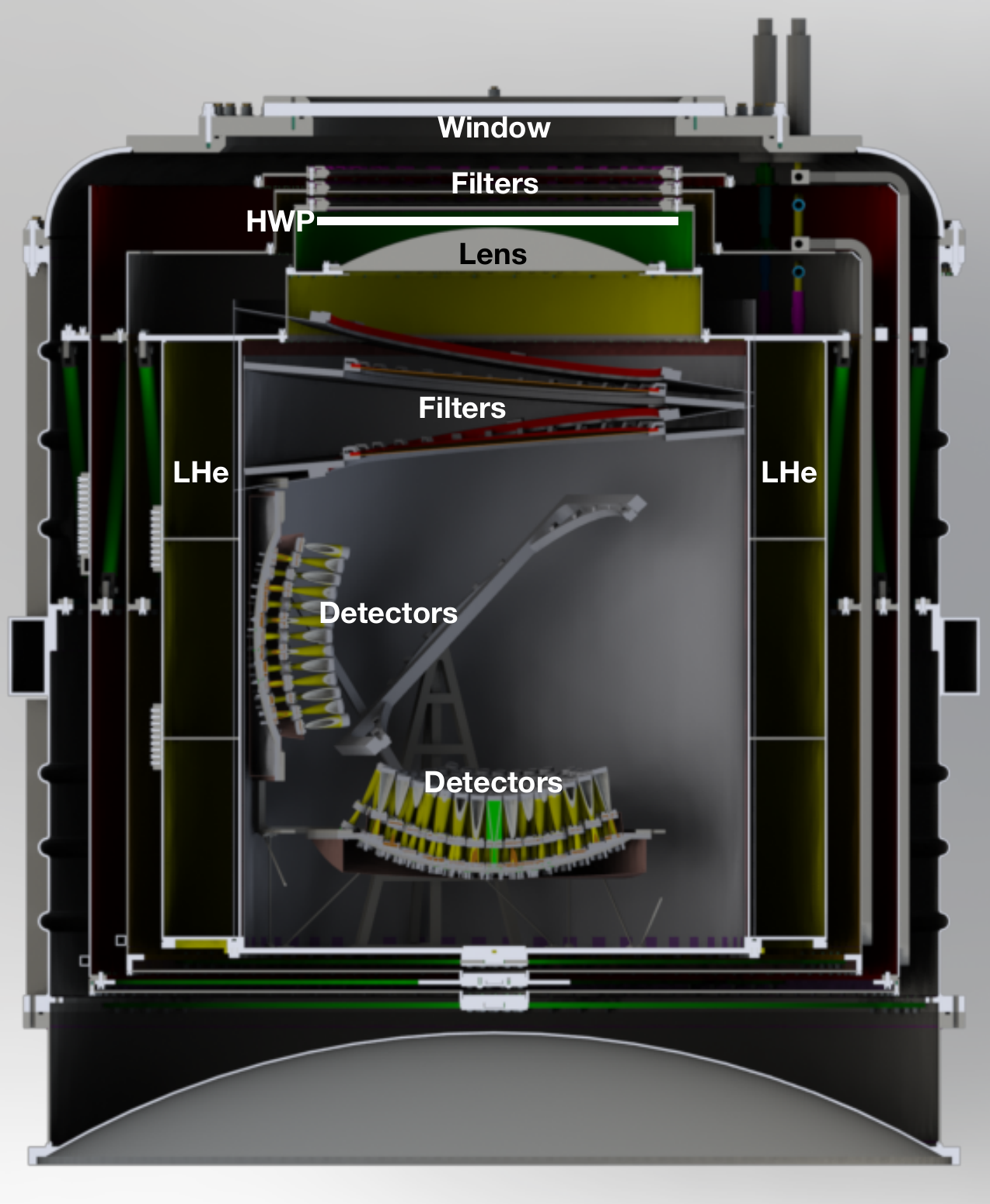}}
  \caption{A section of the SWIPE cryostat and polarimeter. The main components are labelled in the figure.}
\label{fig:cryosec}
\end{figure}

The SWIPE optical system consists in a Stokes polarimeter, cooled cryogenically in a large aluminum cryostat suitable for balloon borne photometry \citep{Masi:article} to enhance its performance. 
A stiff structure made of fiberglass tubes supports the toroidal $^4$He tank, which is surrounded by two vapor cooled shields, at \SI{170}{\kelvin} and \SI{40}{\kelvin}. The total volume of the L$^4$He tank is approximately 250 liters, the outer diameter of the system is approximately $\ge$\SI{140}{\centi\meter}, and the height is $\ge$\SI{160}{\centi\meter}. A $^3$He sorption fridge is used to cool-down the detectors arrays at a temperature of \SI{0.3}{\kelvin} (Fig.~\ref{fig:cryosec}).

A sequence of three thermal mesh filters at \SI{170}{\kelvin}, \SI{40}{\kelvin} and \SI{4}{\kelvin}, above the lens, reflects away the large optical load from the cryostat window and the sky. The rotating HWP (500 mm diameter) is the next optical component, followed by a polyethilene lens (500mm diameter, 800mm focal length). The converging beam encounters a large photo-litographed wire-grid (\SI{500}{\milli\meter} in diameter) which acts as a polarization analyzer. The wire-grid is titled at \SI{45}{\degree}, and splits the converging beam into two curved multi-frequency focal planes.

The baseline detectors for the SWIPE instrument are spider-web bolometers with Transition Edge Sensor (TES) thermistors~\citep{Gualtieri:article} and a large absorber area to couple efficiently to a multi-mode beam. The spider-web absorber was designed in order to minimize its cross section for incident cosmic rays present in the stratosphere, which are potentially very dangerous for CMB measurements.

The horns designed for the SWIPE instrument couple efficiently with radiation propagating in free space, and the dimensions of the circular waveguide allow the propagation of multi-modes beams of radiation \citep{lamagna:article}. Moreover they ensure high suppression of stray radiation, through the incoherent superposition of the propagated modes at large angles. Each focal plane hosts 163 horns, for a total of 326 multi-mode pixels divided in the 3 frequency bands: \SI{140}{\giga\hertz} (width 30$\%$), \SI{220}{\giga\hertz} (width 5$\%$) and \SI{240}{\giga\hertz} (width 5$\%$), respectively.

\subsection{Polarization Modulation}

In order to estimate the Stokes parameters Q and U with a single detector, in presence of an overwhelming unpolarized background and important $1/f$ noise the polarized signal has to be modulated. The sky rotates by several degrees during a long-duration polar flight with an observing strategy similar to SWIPE (i.e. full azimuth scan at constant elevation, at $\sim$\SI{2}{rpm}): this provides a first level of modulation. A high-pass filter at \SI{30}{\milli\hertz} can applied to remove slow drifts and part of the $1/f$. Due to the slow modulation, however, significant the low-frequency noise remains. Moreover, turning a measurement of Q completely into a measurement of U requires a sky rotation of \SI{45}{\degree}, which cannot be achieved with sky rotation alone. 

So there is room to improve significantly by adopting a further level of modulation. Rotating the entire instrument about its optical axis is one way to allow the same detector to measure both Q and U. However, while the desired polarization sensitivity rotates with the instrument, many undesired instrumental systematic effects arise. To improve the filter efficiency the simplest way is to rotate one element of the Stoke polarimeter with respect to the rest. If we chose to rotate the polarizer at a frequency $f$ the intensity measured by the detector will be modulated at twice the mechanical frequency, while if we rotate the HWP with respect to the polarizer we obtain:
\begin{equation}
I(t) = \frac{1}{2}(T + Q \cos 4\omega t + U \sin 4\omega t)
\end{equation}

The HWP can be stepped or spun. The former is easier to implement, but produces a slow modulation. The latter is more difficult to implement due to the constraints on the dissipated power in the cryogenic system, but a mechanical spin frequency of $\sim$ \SI{1}{\hertz} allows to move the signal of interest to $\sim$ \SI{4}{\hertz}. If the modulation frequency is well above the $1/f$ knee of the noise power spectrum, white-noise sensitivity can be achieved on long timescales in polarization. Continuous HWP rotation for polarization modulation is thus very efficient to suppress $1/f$ noise. The sky data is extracted via demodulation during data analysis, knowing the position of the HWP with the same sampling rate of the detectors. 

A first analysis of the effectiveness of this strategy in the case of SWIPE has been described in \citep{Buzzelli2018:article}. Here we perform an independent analysis, focusing on the efficiency in the removal of atmospheric and instrumental $1/f$ noise and other systematic effects. 

In order to test the impact of the HWP on CMB polarization measurements by SWIPE we use a code able to generate a realistic SWIPE scanning strategy in presence of spinning or stepped HWP. The main parameters are reported in Tab.~\ref{tab:params}. We complete this software with a map-making algorithm which collapses data timelines into maps. 

\begin{table}[h!]
\centering
\begin{tabular}{lccc} 
 Strategy & HWP      & Payload  & Mission  \\[0.4ex]  
          & rotation & rotation & duration \\[0.4ex]  
 {} & [\SI{}{\hertz}] & [RPM] & [days] \\[0.4ex]
 \hline
Step & 1.4 10$^{-3}$ & 2.0 & 14 \\
Spin & 1.0 & 2.0 & 14\\[0.4ex]
\end{tabular}
\caption{Mission parameters for the scanning strategies adopted in the simulation pipeline.}
\label{tab:params}
\end{table}

We performed noise-free simulations of the observation of the CMB sky through the SWIPE scan strategy, to understand the effect of the application of a high-pass filter on the measured data to reconstruct the input sky map. The high-pass filter in fact is used to remove the contribution of $1/f$ noise, but also removes a fraction of the CMB sky signal. 

\begin{figure}[ht]
\centering
{\includegraphics[scale=0.43]{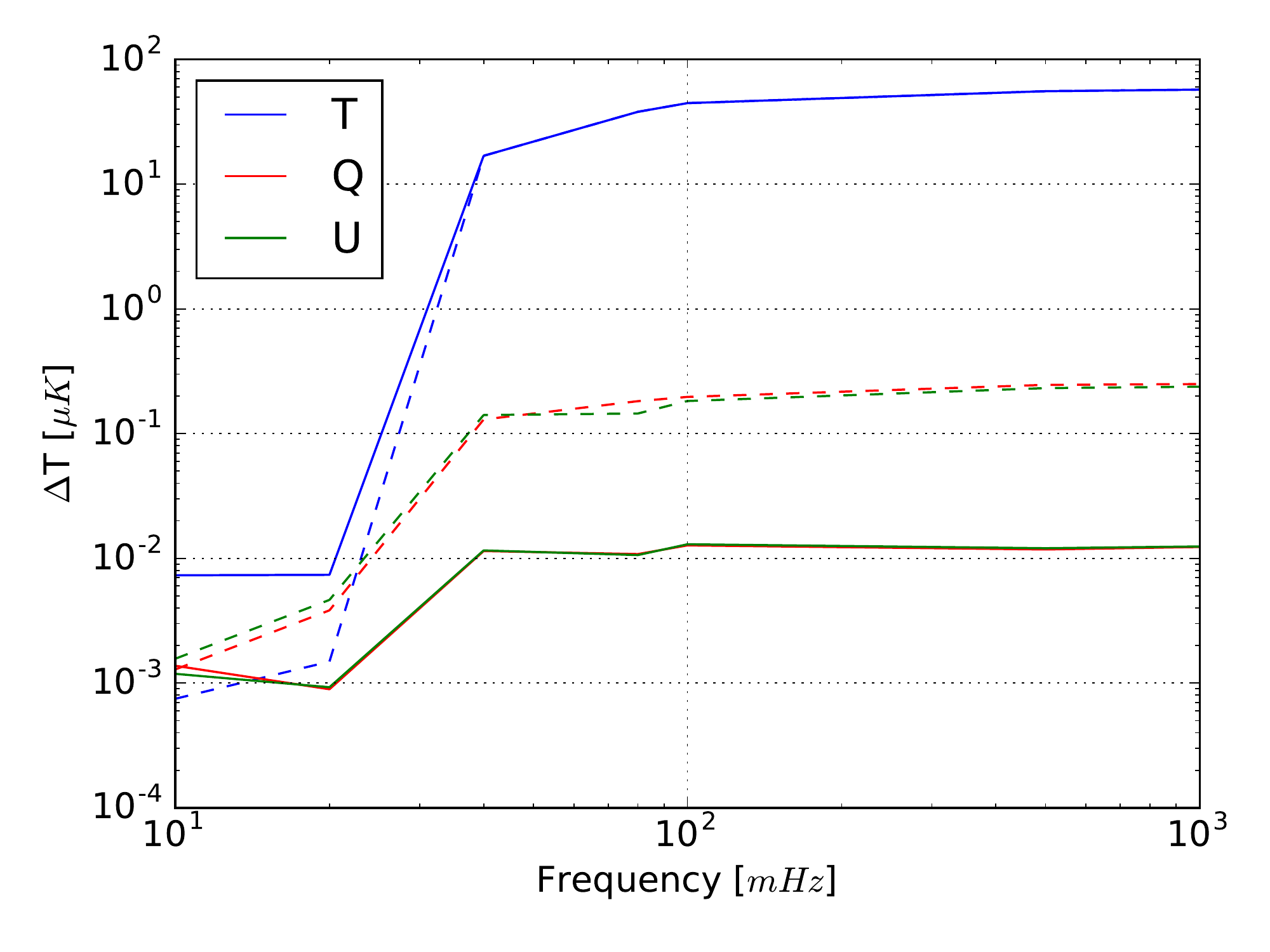}}
  \caption{Standard deviations of the difference between the output and the input maps for T, Q and U. The smaller the standard deviation, the better the quality of the reconstruction. Solid lines correspond to spinning HWP while dashed lines correspond to stepped HWP.}
\label{fig:all}
\end{figure}

As a first indicator, we compute the standard deviation of the difference between the output and the input maps. This provides an estimate of the quality of the reconstructed map. Fig.~\ref{fig:all} shows results as a function of the cut-on frequency of the high-pass filter (filter slope =\SI{0.1}{\milli\hertz}$^{-1}$). The stepped HWP case is represented by dotted lines while solid lines represent the continuously spinning HWP case.

Lower values of the standard deviation correspond to a better reconstruction of the input map. We find that cut-on frequencies higher than 10-\SI{20}{\milli\hertz} result in significant differences between the original and the reconstructed maps. The Q and U maps are reconstructed better for the spin case, where the standard deviation of the difference map is smaller by one order of magnitude with respect to the stepped case, for cut-on frequencies > 30 mHz. 

Since the stepped rotation produces a modulated signal at frequencies < 10 mHz, the result above is already a strong indication in favour of the continuous spinning case. 

The standard deviation indicator does not provide any information about the angular scales which are reconstructed better or worse. As a second, more informative indicator, we calculated the angular power spectra $C_l$.

\begin{figure}
\subfloat[(a) TT power spectra.]{\includegraphics[width = 3.3in]{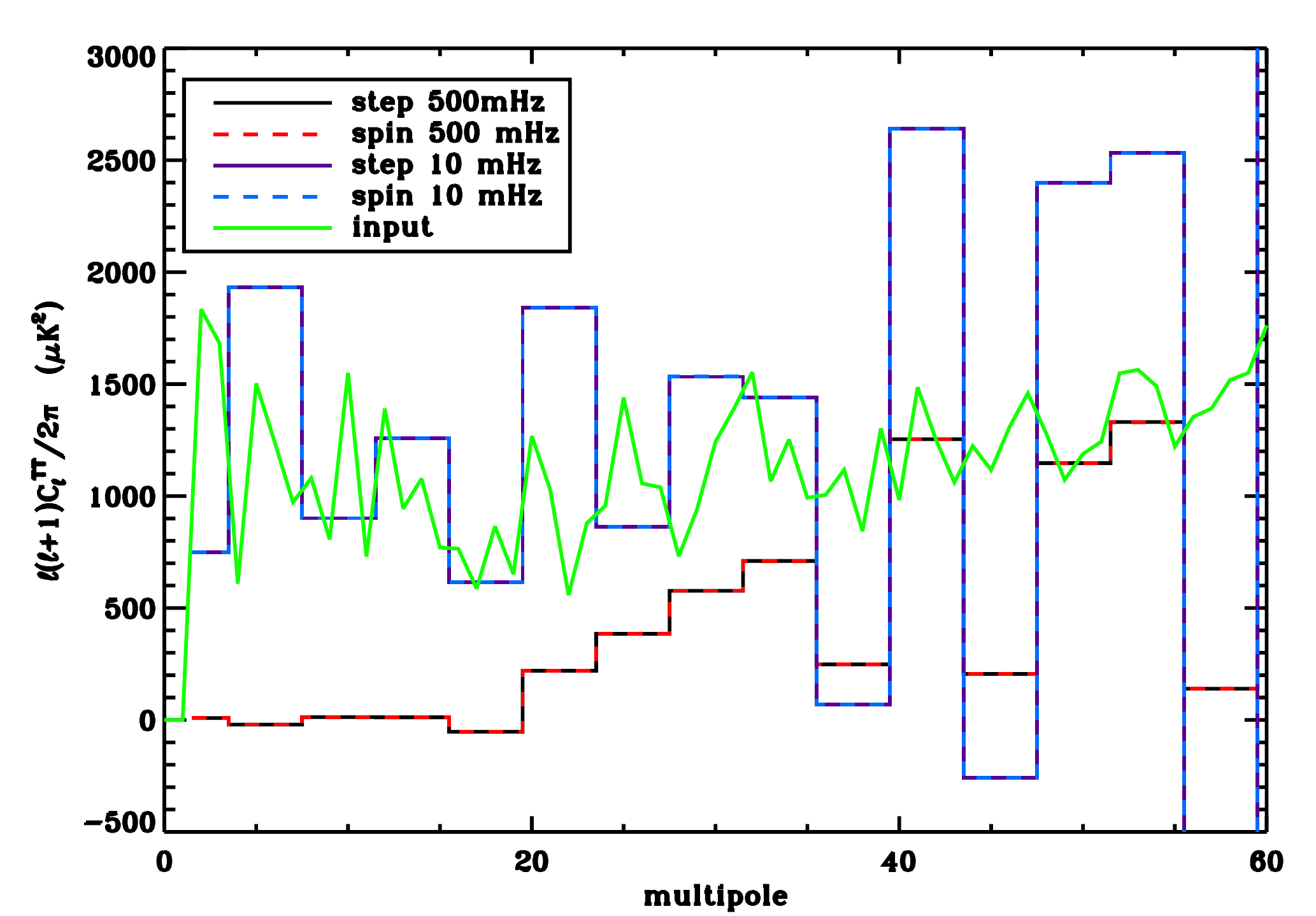}} \\
\subfloat[(b) EE power spectra.]{\includegraphics[width = 3.3in]{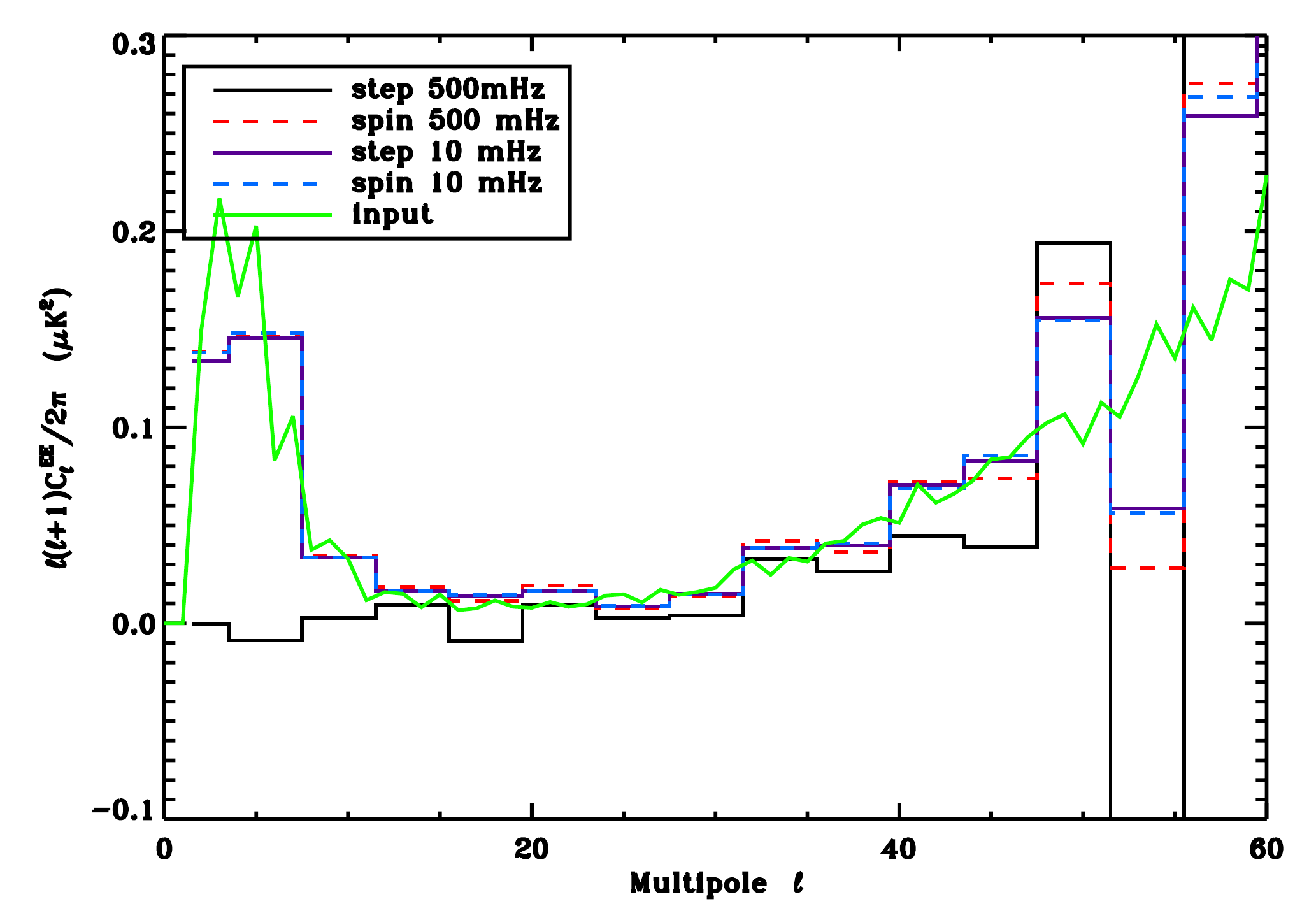}}
\caption{Green lines represent the input power spectrum for TT and EE. (a) TT power spectra reconstructed better with lower cut frequency both for spin and step cases. (b) EE power spectra reconstructed well for a spin HWP independently of the cut frequency, while the step case works properly only for low cut frequency.}
\label{fig:cl}
\end{figure}

We analyzed two extreme cases in opposite directions, with cut-ons at \SI{10}{\milli\hertz} and at \SI{500}{\milli\hertz}. Panel (a) of Fig.~\ref{fig:cl} shows the reconstructed temperature power spectra (TT), compared to the power spectrum of the input map (green). We find that the TT signal is reconstructed well only for the lower cut frequency case, for both strategies. Panel (b) of Fig.~\ref{fig:cl} shows the E-modes power spectra (EE): we find that with a step HWP we are able to reconstruct the spectrum if and only if the cut frequency is very low, while for the spinning case we can reconstruct the polarized signal regardless of the cut frequency. 

Figures ~\ref{fig:all} and ~\ref{fig:cl} refer to unrealistic noiseless measurements. In a realistic noise scenario the most important contribution comes from $1/f$ noise, extremely high at lower frequencies. As we have just seen, with a stepped HWP configuration we have to set the cut frequency $f_c<$\SI{20}{\milli\hertz}. This means that if the knee frequency of the $1/f$ noise is comparable or higher respect to $f_c$, a lot of noise will contaminate our data. 
We conclude that a continuously spinning HWP is better suited to this kind of experiment, allowing to filter out very efficiently $1/f$ noise. In fact, in the continuously spinning HWP case, the cosmological signal is contained in a narrow band around $4f_{spin}$, allowing the use of a band-pass filter to reject all the noise outside this band.

\section{HWP properties}
\subsection{Temperature}
While a continuous rotating HWP is the best solution from the point of view of $1/f$ noise rejection, the choice of the operating temperature is not trivial.  The spurious signals produced by the HWP, due to non-idealities in the HWP itself and in the optical system of the polarimeter, is due at least in part to the emissions of the HWP and of the polarizer, which decrease with their temperature. However, the lower the temperature, the weaker the cooling power of the cryogenic system. So we need a trade-off, based on quantitative estimates.

\begin{figure}[ht]
\centering
{\includegraphics[scale=0.43]{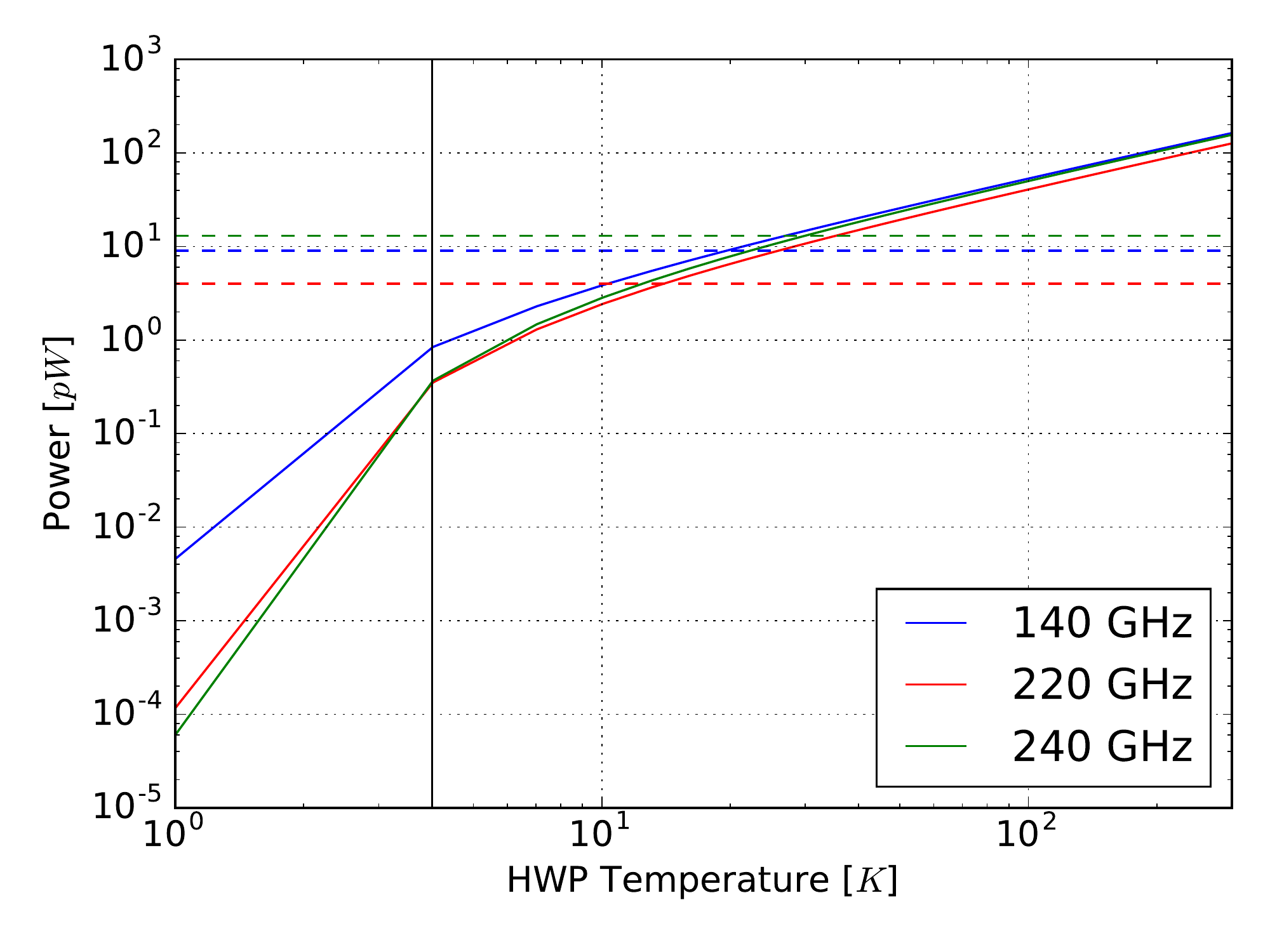}}
  \caption{Power load produced by the HWP for each band as a function of the HWP temperature (color dashed lines). The dashed lines correspond to the background power from the sky and the rest of the optical system. The vertical black line corresponds to the maximum temperature ($\sim$\SI{4}{\kelvin}) where the HWP contribution is negligible.}
\label{fig:power_hwp}
\end{figure}

Each detector sees the HWP as a grey-body with an emissivity $\epsilon_{HWP}(\nu)$. The power load on the focal plane is:
\begin{equation}
P(\nu, T) = \int B(\nu, T) \, \epsilon_{HWP}(\nu) \, \epsilon_f(\nu) \, A\Omega \, d\nu
\end{equation}
where $B(\nu,T)$ is the blackbody spectrum and $\epsilon_f(\nu)$ the transmission of the band-pass filters mounted on each detector (Fig.~\ref{fig:power_hwp}). Fig.~\ref{fig:power} shows the power load produced by the HWP for each band as a function of the HWP temperature. The color dashed lines correspond to the background power produced by the two main contributions: the sky and the window (\SI{9}{\pico\watt} at \SI{140}{\giga\hertz}, \SI{4}{\pico\watt} at \SI{220}{\giga\hertz} and \SI{13}{\pico\watt} at \SI{240}{\giga\hertz}). The contribution of the HWP emission on the total power load should be as small as possible. At $\sim$\SI{10}{\kelvin} the power load produced by the HWP is comparable to the background power, while at \SI{4}{\kelvin} or less the contribution becomes negligible.

\subsection{Spurious signals}

The most important spurious signals, in a Stokes polarimeter of this kind, comes from polarized signals produced internally which can be confused with the cosmological signal modulated at $4f$~\citep{Salatino:article}. For a tensor to scalar ratio $r = 0.1$, the expected detected signal due to the CMB only, has a typical (rms) amplitude of about \SI{0.7}{\micro\kelvin}. To convert the power emitted by each optical element into equivalent CMB fluctuations we use the following relation:
\begin{equation}
\Delta T = \frac{\int B(\nu, T) \, \epsilon_f(\nu) \, A\Omega \, d\nu}{\int \frac{\partial B(\nu, T)}{\partial T} \, \epsilon_f(\nu) \, A\Omega \, d\nu}
\end{equation}

Small differences in the absorption coefficient ($\sim 10^{-3}$) of the HWP produce a polarized emission; this radiation is modulated at $2f$ when is transmitted by the polarizer. This radiation could also be reflected by the polarizer, successively by the HWP and modulated at $4f$, the same frequency of the cosmological signal.

\begin{figure}[ht]
\centering
{\includegraphics[scale=0.43]{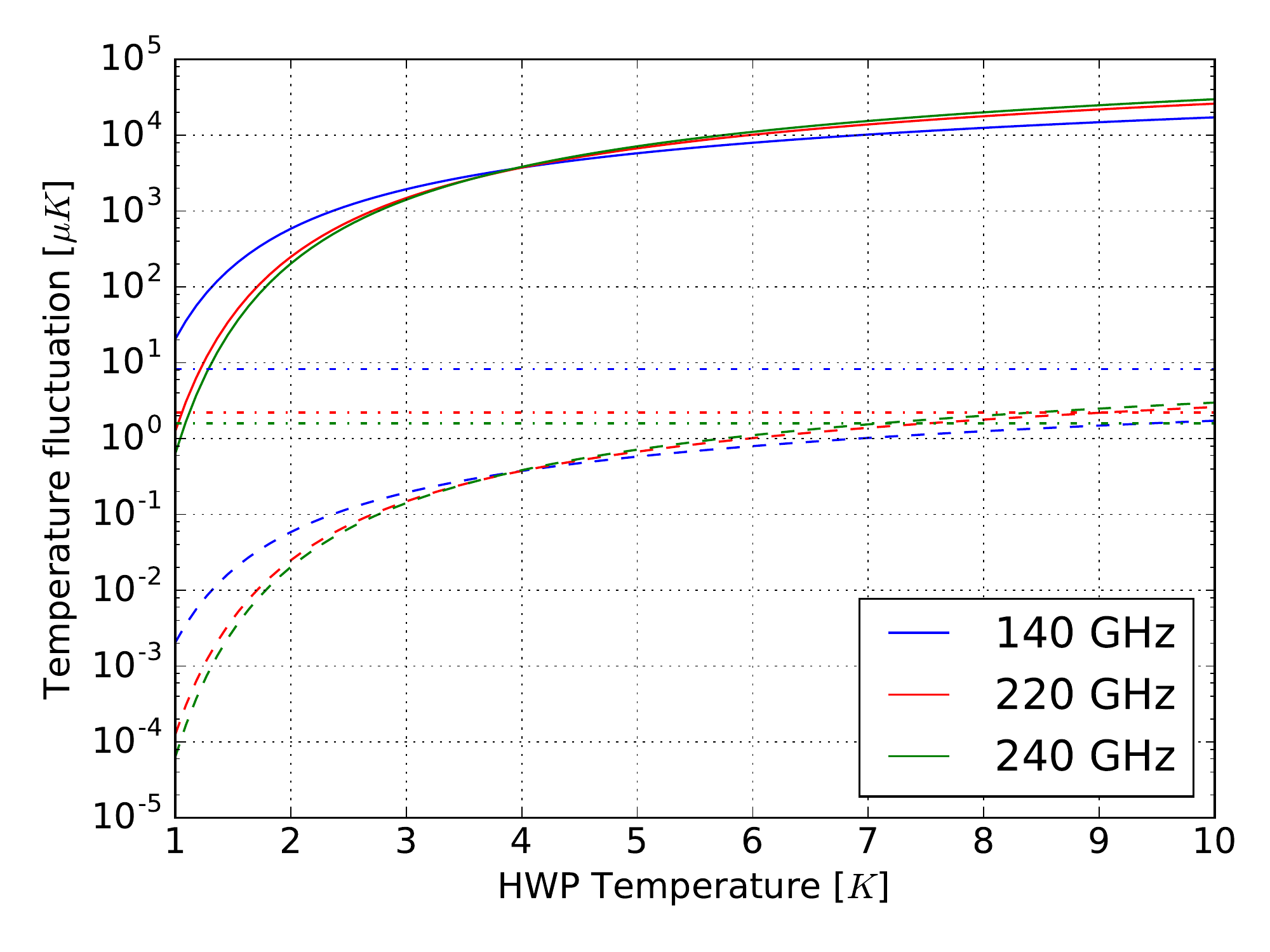}}
  \caption{Solid lines: polarized emission of the HWP transmitted by the polarizer. Dashed lines: polarized emission of the HWP reflected by the polarizer and reflected again by the HWP. Dash-dot lines: polarized emission of the polarizer at \SI{1.6}{\kelvin} reflected by the HWP.}
\label{fig:power}
\end{figure}

Fig.~\ref{fig:power} shows the expected signals modulated at $2f$ (solid lines) and $4f$ (dashed lines). The huge $2f$ contribution could be removed by using a high pass filter with a cut frequency between $2f$ and $4f$ to reject the entire spurious signal. On the other hand the $4f$ contribution could not be removed but only characterized carefully or minimized by reducing and maintain constant the HWP temperature (the polarization modulator of SWIPE is cooled at \SI{1.6}{\kelvin}).

Another critical element is the polarizer, because its linear polarized emission is reflected back by the rotating HWP, and transmitted by the polarizer itself, so that it is modulated at $4f$. If we assume \SI{1.6}{\kelvin} as the temperature of the polarizer (as in SWIPE), the signals of all three bands correspond to few \SI{}{\micro\kelvin}, the same order of magnitude of the cosmological signal. Instead of the previous $4f$ signal, this one is simpler to characterize because the HWP reflectivity is basically independent of the temperature. Otherwise this spurious signal could be reduced dramatically by cooling down the polarizer below \SI{1}{\kelvin} (i.e. connecting it at \SI{0.3}{\kelvin}. This is not always possible due to the limited cooling power of sub-K fridges).



\section{Conclusions}\label{sec5}

In this paper we discussed the importance of using a continuously rotating HWP in the Stokes polarimeter of the SWIPE experiment. We generated a realistic SWIPE scanning strategy in presence of spinning or stepped HWPs, and evaluated the ability of reconstructing the input sky map simulating the CMB. 
The ability of spinning the HWP to reconstruct the input map, despite of the cut frequency of the applied high pass filter, was demonstrated. This drove the choice to this configuration for SWIPE. So we modulate the cosmological signal at 4 times the mechanical rotation frequency and move it far away from the $1/f$ noise, which is easily filtered out.

The temperature of the HWP (1.6K in SWIPE) was selected to minimize the background power on the detector and the effect of spurious signals which arise at $2f$ and $4f$. The magnitude of the latter is also related to the temperature of the polarizer (\SI{1.6}{\kelvin} in SWIPE) and has to be characterized carefully to remove it from cosmological sky data, since is of the same order of magnitude. Note, however, that this spurious signal is not synchronous with the scan of the sky, and has to be considered more like an additional source of noise in the measurements.


\section*{Acknowledgments}

This research has been supported by the Ph.D. fellowship of F.C. and by the "Avvio alla ricerca" grant of the Sapienza University of Rome.



\noindent

\nocite{*}
\bibliographystyle{Wiley-ASNA}
\bibliography{Bib}%

\end{document}